\documentclass[11pt]{article}

\usepackage{amsmath}
\usepackage{amssymb}
\usepackage{amsthm}
\usepackage{graphicx,epsfig}

\newcommand{\etal}{\emph{et al.}\,}

\newcommand {\E}{\mathbf E}
\newcommand {\T}{\mathcal T}

\newcommand{\lpopt}{\gamma^*}

\newcommand {\A}{\mathcal A}

\newcommand{\p}{\mathbf{p}}
\newtheorem{theorem}{Theorem}[section] 
\newtheorem{lemma}[theorem]{Lemma}
\newtheorem{claim}[theorem]{Claim}

\newtheorem{ex}{Example}[section]

\renewcommand{\paragraph}[1]{\medskip \noindent {\bf{#1}}}

\newcommand{\R}{\mathcal{R}}

\newcommand{\eat}[1]{}
\advance \topmargin by -\headheight
\advance \topmargin by -\headsep
\evensidemargin \oddsidemargin
\renewcommand{\O}{\mathcal{O}}
\newcommand{\G}{\mathcal{G}}

\begin{document}
\title{ Sequential Design of Experiments via Linear Programming\thanks{This
    combines work from two papers~\cite{GuhaM07,GuhaM07b} appearing in the
    ACM-SIAM Symposium on Discrete Algorithms (SODA), 2007 and the
    $39^{th}$ Annual ACM Symposium on Theory of Computing (STOC), 2007
    respectively.}}  \author{Sudipto Guha\thanks{Department of Computer
    and Information Sciences, University of Pennsylvania.  Email: {\tt
      sudipto@cis.upenn.edu}. Research supported in part by an Alfred P.
    Sloan Research Fellowship and by an NSF Award CCF-0430376.}  \and
  Kamesh Munagala\thanks{Department of Computer Science, Duke University.
    Email: {\tt kamesh@cs.duke.edu}. Research supported in part by NSF
    CNS-0540347.}}  \date{}
  \maketitle

\newcommand{\Th}{\Theta}
\renewcommand{\S}{\mathcal{S}}
\renewcommand{\H}{\mathcal{S}}
\renewcommand{\T}{\mathcal{S}}
\newcommand{\N}{N_{\epsilon}}

\thispagestyle{empty}

\begin{abstract}
  The celebrated multi-armed bandit problem in decision theory models the central trade-off  between exploration, or learning about the state of a system, and exploitation, or utilizing the system. In this paper we study the  variant of the multi-armed bandit problem where the exploration phase involves costly experiments and occurs before the exploitation phase; and where each play of an arm during the exploration phase updates a prior belief about the arm. The problem of finding an inexpensive exploration strategy to optimize a certain exploitation objective  is {\sc NP-Hard} even when a single play  reveals all information about an arm, and all exploration steps cost the same. 
  
  We provide the first polynomial time constant-factor approximation algorithm for this class of problems.  We show that this framework also generalizes several problems of interest studied in the context of data acquisition in sensor networks. Our analyses also extends to switching and setup costs, and to concave utility objectives.

Our solution approach is  via a novel linear  program rounding technique based on stochastic packing. In addition to yielding exploration policies whose performance is within a small constant factor of the adaptive optimal policy, a nice feature of this approach is that the resulting policies explore the arms {\em sequentially} without revisiting any arm.  Sequentiality is a well-studied paradigm in decision theory, and is very desirable in  domains where multiple explorations can be conducted in parallel, for instance, in the sensor network context.
\end{abstract}

\section{Introduction}
The sequential design of experiments is a classic problem first formulated
by Wald in 1947~\cite{Wald47}. The study of this problem gave rise to the
general field of decision theory; and more specifically, led
Robbins~\cite{Robbins} to formulate the celebrated multi-armed bandit
problem, and Snell~\cite{Snell} and Robbins~\cite{Robbins} to invent the
theory of optimal stopping. The copious literature in this field is
surveyed by Whittle~\cite{W82,W83}.

The canonical problem of sequential design of experiments is best
described in the language of the multi-armed bandit problem: There are $n$
competing options referred to as ``arms'' (for instance, consider clinical
treatments) yielding unknown rewards (or having unknown effectiveness)
$\{p_i\}$. Playing an arm (or testing a treatment on a patient) yields
observations that reveal information about the underlying reward or
effectiveness. The goal is to sequentially test the treatments (or
sequentially play the arms) in order to ultimately choose the ``best''
one.  Such problems are usually studied in a decision theoretic setting,
where costs and utilities are associated with actions (testing a
treatment) and outcomes (choosing one treatment finally). The goal of any
decision procedure is to come up with a plan for testing the treatments
(or playing the arms) and choosing an outcome in order to optimize some criterion based on the
costs and utilities. The testing procedure is termed {\em exploration}, and
choosing the outcome is termed {\em exploitation}. The crux of the
multi-armed bandit problem, and the reason has been extensively studied,
is that it cleanly models the general trade-off between the cost of
exploration (or learning more about the state of the system) and the
utility gained from exploitation (or utilizing the system).

\paragraph{}
Various frameworks in decision theory differ in (i) the available
information and (ii) optimization criteria for evaluating a decision plan.
We now describe the problem we study from the perspective of these design
choices. From the perspective of available information, we focus
exclusively on the Bayesian setting, first formulated by Arrow, Blackwell
and Girshick in 1949~\cite{ABG49}.  In this setting, each arm (or
treatment) is associated with prior information (specified by
distributions) that updates via Bayes' rule conditioned on the results of
the plays (or tests). More formally, we are given a bandit with $n$
independent arms. The set of possible states of arm $i$ is denoted by $\S_i$, and the initial state is $\rho_i \in \S_i$.  When the arm $i$ is played in a state $u \in \S_i$, the arm transitions to state $v \in \S_i$ w.p. $\p_{uv}$ depending on the observed outcome of the play.  The initial state models the prior knowledge about the arm. The states in general capture the posterior conditioned on the observations from a sequence of plays (or experiments) starting at the root.  The cost of a play depends on whether the
previous play was for the same arm or not. If the previous play was for
the same arm, the play at $u \in \S_i$ costs $c_u$, else it costs $c_u +
h_i$, where $h_i$ is the setup cost for switching into arm $i$\footnote{Our algorithms also extend to concave costs where the cost of $r$ consecutive play as
  well as switching out costs, we omit that discussion here.}.  Recall
that the arms correspond to different treatments or experiments;
therefore, this cost models setting up the corresponding experiment. Every state $u \in \S_i$ is associated with a reward $r_u$, which is the expected reward of playing in this state (which is of course conditioned on the observations from the plays so far).  By Bayes' rule, the reward of the different states evolve according to a {\em Martingale} property: $r_u = \sum_{v \in \S_i} \p_{uv} r_v$. We present concrete examples of state spaces in Section~\ref{sec:2level}.

\paragraph{}
From the optimization perspective, our objective is to maximize {\em
  future utilization}. Any policy explores (or tests) the arms for a
certain amount of time and subsequently, exploits (or chooses) an arm that
yields the best expected posterior (or future) reward. For this objective
to be meaningful, we need to constrain the total cost we can incur in
exploration before making the exploit decision.  A natural example of this
is product marketing research, where the {\em entire exploration phase
  appears before the exploitation phase}. Formally, a policy $\pi$
performs a possibly adaptive sequence of plays during the exploration.
Since the state evolutions are stochastic, the exploration phase leads to
a probability distribution over outcomes, $\O(\pi)$.  In outcome $o \in
\O(\pi)$, each arm $i$ is in some final state $u_i^o$. In this outcome $o$
the policy will choose the ``best arm'' $\max_i r_{u^o_i}$ (or a suitable
concave function of the vector $\langle \cdots, r_{u^o_i}, \cdots
\rangle$). The expected reward of the policy $\pi$ over the outcomes of
exploration, $R(\pi)$ is $\sum_{o \in \O(\pi)} q(o,\pi) \max_i r_{u^o_i}$.
Let $C(o,\pi)$ denote the cost of the exploration plays made by the policy
given an outcome $o$. In the simplest version, we seek to find the policy
$\pi$ which maximizes $\R(\pi)$ subject to $C(o,\pi) \leq C$ for all $o
\in \O$. As remarked in~\cite{ABG49}, this problem is solvable by dynamic
programming~\cite{Bellman57,book}.  However this approach requires
computation time polynomial in the joint state space (truncated by the
budget constraint) for multiple arms, which is the {\bf product} of the
individual (truncated) state spaces. Unsurprisingly, the problem becomes
{\sc NP-Hard} even when a single play reveals the full information about
an arm, and all plays (across different arms) cost the
same~\cite{GoelGM06}. Designing a policy which is computationally
tractable, at the cost of bounded loss in performance, is the main goal of
this paper. We will study the problem from the perspective of
approximation algorithms, where we seek to find a provably near optimal
solution with the restriction that the algorithm must run in time
polynomial in the sum of the state spaces. More precisely, we seek an
algorithm which would give us an utilization least $OPT/\alpha$ where
$OPT=\max_\pi \R(\pi)$ subject to $C(o,\pi) \leq C$ for all $o \in \O$;
which is denoted as an $\alpha$ approximation. Note that we seek a
multiplicative approximation because such a result is invariant under
scaling of the rewards (see also the discussion on discount rewards
below). Since it is {\sc NP-Hard} to determine $OPT$, we seek to use a
linear program to determine an upper bound $\gamma^* \geq OPT$ and provide
an algorithm that achieves $\gamma^*/\alpha$ in the worst case. The added
benefit of such an approach is that we have a concrete upper bound
$\gamma^*$ for comparison and an algorithm which guarantees
$\gamma^*/\alpha$ in the worst case, may have a significantly better (and
quantifiable, due to the existence of the upper bound) performance in
practice. 
The interested reader may consult
\cite{vazirani01} for a review of approximation algorithms.

The necessity of studying this problem is further hastened by the
emergence of several applications where the number of arms is large,
typically data intensive applications. Examples of this problem arise in
``active learning'' \cite{madani,moore} where the goal is to learn and
choose the most discerning hypothesis by sequentially testing the
hypotheses on a set of assisted examples; sensor networks~\cite{Krause},
where the goal is sensor placement to maximize a utility function such as
information gain, based on sequentially collecting a small number of
samples; and databases~\cite{Shiv05}, where the goal is to settle upon a
possibly long running query execution plan, again based on a few carefully
chosen samples.

\subsection{Related Models}
The future utilization objective is well-known in literature (refer for
instance, Berry and Fristedt~\cite{Berry}, Chapter $3.6$). The unit cost
version of this problem is a special case of the {\em infinite horizon
  discounted} multi-armed bandit problem. In the discounted bandit
problem, there is an infinite discount sequence $\{\alpha_t \in [0,1]| t =
1,2, \ldots\}$. Any policy $\pi$ plays an arm at each time step; suppose
the expected reward from playing at time $t$ is $R_t(\pi)$. The goal is to
design an adaptive policy $\pi$ to maximize $\sum_{t \ge 1} \alpha_t
R_t(\pi)$. The future utilization objective with an exploration budget $C$
corresponds to $\alpha_1 = \alpha_2 = \cdots = \alpha_{C} = \alpha_{C+2} =
\alpha_{C+3} = \cdots = 0$, and $\alpha_{C+1} = 1$. This setting implies
the objective is the reward of the arm chosen at the $(C+1)^{st}$ play
(exploitation), and only plays of significance for making this choice are
the first $C$ plays (exploration). As observed in~\cite{Berry}, this
problem seems significantly harder computationally than the case where the
discount sequence is monotonically decreasing with time.  In fact, when
the discount sequence is geometric, {\em i.e.}, $\alpha_t = \beta^t$ for
some $\beta < 1$, the celebrated result of Gittins and Jones shows that
there exists an elegant greedy optimal solution termed the {\em Gittins
  index} policy~\cite{GJ74}; an index policy ranks the arms based solely
on their own characteristics and plays the best arm at every step.  The
Gittins index is suboptimal both the finite horizon setting where
$\alpha_t = 1$ for $t \le C$ and $0$ otherwise; as well as the
future utilization setting we consider here~\cite{madani}. Finally, Banks
and Sundaram \cite{BS94} show that no index exists in the presence of
switching in/out costs.

Alternatives to the Bayesian formulation are also as old as the original
study of Wald~\cite{Wald47} and Robbins~\cite{Robbins}. These versions do
not assume prior information, but instead perform a min-max optimization
over possible underlying rewards via a suitably constructed loss or regret
measure.
As observed in~\cite{Berry,WG}, although minmax objectives are more
robust, the Bayesian approach is more widely used since it typically
requires less samples.  Furthermore, the regret criterion naturally forces
the optimization to consider the {\em past}: What is the minimum loss in
the past $N$ trials due to not knowing the true rewards.  Note that
minimizing regret is not the same as maximizing future utilization, the
former being more akin to the finite horizon version with discount
sequence $\alpha_t = 1$ for $t \le C$ and $0$ otherwise. Intuitively, in
the former, we attempt to minimize the error {\em during} the testing
process, while in the latter, we do not care about errors in testing, but
attempt to ensure that at the end, we are truly picking the (near) best
option for exploitation.

Nevertheless, it is natural to ask whether the algorithms suggested in the
context of minmax analysis, particularly the seminal works of Lai and
Robbins~\cite{Lai}, and Auer, Cesa-Bianchi and Fischer \cite{Auer3} (and
extended to uniform switching costs in~\cite{VPT92,Auer2}), have good
performance guarantees in the future utilization measure. However these
are ``model free'' algorithms, and it is easy to show that for
appropriately chosen budget $C$, these algorithms have significantly
inferior performance on the future utilization objective as compared to
algorithms that use the prior information. This is not surprising because
the objectives are different. Similar comments apply to the
``experts'' problem \cite{experts} and subsequent research in adversarial
multiarmed bandits \cite{Auer,FlaxmanKM05} where the reward distribution
is chosen by an adversary and need not be stochastic.

It is worth pointing out that in the loss function or minmax approach, the
loss or regret arises due to lack of information about the rewards. The
difficulty in optimizing future utilization in the Bayesian setting arises
from the computational aspect. This is quite similar to the differences
between the classes of online and approximation algorithms.

\subsection{Structure of the Policies}
For the future utilization measure, it is worth mentioning that the
general structure of the policies are important. Two such classes of
policies are noteworthy. The first class is motivated by the stopping time
problem, an early example of which is the secretary problem~\cite{CMRS64}. A policy in this class {\em fixes an ordering of the arms in advance}, and samples the arms sequentially, {\em i.e.},  does not return to
previously rejected arm. The benefit of such strategy is
that these are often succinct to represent and easy to implement in real
hardware from the perspective of control. Another benefit, as the reader
would have observed, is that it is easy to model switching/setup costs in
such policies; these costs in fact can be generalized so that $r$
consecutive plays have a cost which is concave function in $r$.  We define
such policies as {\bf sequential}, because the ordering of the arms is fixed
beforehand. Such strategies have been considered in testing between two
hypothesis~\cite{Wald47}, stochastic
scheduling~\cite{MohringSU,SkutellaU}, stochastic
packing~\cite{DeanGV04,DeanGV05} and in operator placement in databases
\cite{BabuMMNW04,BabuMWM05} -- however all except the hypotheses testing
results hold for {\bf two-level} state spaces (or arms with point priors),
where a single play reveals complete information about the underlying
reward of the arm. (Refer Section~\ref{sec:2level} for a formal definition.)

The second and more restrictive class of policies performs all the tests
(or plays) before observing any of their outcomes. Therefore, the policy
has three disjoint successive phases: Test, observe, and select.  Such
{\bf non-adaptive} policies are of interest when the observations can be
made in parallel, and therefore the final choice can be made quicker.
Naturally these strategies are meaningful for two level state spaces, and
have thus been found to be of interest in context of sensor
networks~\cite{Krause}, multihoming networks~\cite{AkellaMSSS03},
stochastic optimization~\cite{GoelGM06,GuhaM07} and database
optimization~\cite{Shiv05}.

For both the above classes, the {\em goal} is to show that performance of
an algorithm that is restricted to the respective class is not
significantly worse compared to an adversary whose strategy is {\em fully
  adaptive}. This is known as the {\em Adaptivity Gap} of a strategy. All
previous analysis of adaptivity gap was restricted to two level state
spaces. This paper provides an uniform framework that extends to both the
classes above and applies to multilevel state spaces. It is interesting to
note that one of the original goals of Wald~\cite{Wald47} in sequential
analysis was to explore sequential strategies. Though such strategies are
optimal for choosing between two hypothesis, the difficulty in obtaining
optimal strategies for testing multiple competing hypotheses was known
since that time. {\em The major contribution of this work is to show that
  in a variety of bandit settings, when we are seeking to optimize any
  concave function of the posterior probabilities, the adaptivity gap in
  considering sequential strategies is bounded by a constant.} In other words,
the performance of a fully adaptive solution cannot be significantly
better than a sequential strategy.

\subsection{Problems and Results}
\label{sec:probs}
We consider three main types of problems in this paper. Recall that there are $n$ independent arms, each with its own state space $\S_i$; a policy $\pi$ adaptively explores the arms paying expected cost $C(\pi)$ before selecting an arm for exploitation based on the observed outcomes. The expected reward of the selected arm over the outcomes of the policy $\pi$ is denoted $R(\pi)$. 

\begin{itemize}
\item {\bf Budgeted (Futuristic) Bandits:} There is a cost budget $C$. A policy $\pi$
  is {\em feasible} if for any sequence of plays made by the policy, the
  cost is at most $C$. The goal is to find the feasible policy $\pi$ with
  maximum $R(\pi)$. We have already discussed {\bf switching costs.}  An
  extension of switching cost is {\bf concave play cost} where the cost of
  sequential interrupted plays of an arm is concave in the number of
  plays. This was first hinted at in~\cite{ABG49} and the authors explicitly
  settled on linear costs.

  A generalization of the above problem is {\bf budgeted concave utility
    bandits} problem where the objective function is an arbitrary concave
  function of the final rewards of the arms. Examples of such
  function include choosing the best $K$ arms, power allocation across
  noisy channels \cite{cover} or optimizing ``TCP friendly'' network
  utility functions ~\cite{low}.

\item {\bf Model Driven Optimization:} This is a non-adaptive formulation
  of the above, where the state space $\S_i$ is $2$-level and a single play
  reveals full information about an arm.  In such a context, {\bf non-adaptive} strategies are
  desirable since the plays can be executed in parallel. A feasible non-adaptive
  policy $\pi$ chooses a subset of the arms to explore, before seeing the
  result of any of the plays. There has been a significant number of
  papers in recent years, specially in the context of sensor networks. Our
  paper unifies this thread with the bandit framework. 

\item {\bf Lagrangean (Futuristic) Bandits:} Find the policy $\pi$ with maximum $R(\pi) -
  C(\pi)$. Note that the Lagrangean can be defined on both the adaptive
  and non-adaptive setting.  This is a natural extension of the single-arm
  optimal stopping time problem.
\end{itemize}

In this paper, we present a single framework that provides efficient
algorithms yielding policies with near-optimal performance for all of the
above problems. For the budgeted (futuristic) bandits in the concave cost
setting (including switching in/out cost), we show that there exists a
sequential strategy that respects the budget, and has objective value at
most a factor $4$ away from that of the optimal fully-adaptive strategy
subjected to the same budget. Section~\ref{sec:2level} discusses different
state spaces. This is presented in
Section~\ref{sec:budget} presents the approximate sequential strategy that
respects the budget, for linear utilities (objective function).
We also present a bicriteria $2(1+\alpha)$
approximation with the cost constraint relaxed by a factor
$\frac{1}{\alpha}$. 
In Section~\ref{sec:nonadapt},
we show how the same framework gives a more restricted
non-adaptive strategy for 2-level states spaces which is within constant
factor of the best adaptive strategy.  In contrast, for multi-level state
spaces, any non-adaptive strategy has a significant performance loss. 
We also present a sequential
strategy that is a $2$ approximation for the Lagrangean Bandits in
Section~\ref{sec:lag}.  In Section~\ref{sec:exten}, we extend the results
in Section~\ref{sec:budget} to concave utilities with a factor $2$ loss of
the approximation factor. 

Note that constant factor approximations are best possible from the
context of  {\em adaptivity gap} of sequential policies as well as {\em integrality gap} of the
linear programming relaxations we use.

\paragraph{Techniques:}  We use a linear programming formulation over the state
space of individual arms, and we achieve polynomial sized formulation in
the size of each individual state space. This particular formulation has
been used in the past~\cite{whittle2,nino} and found to be useful in
practice. To the best of our knowledge, we present the first analysis of
these relaxations in the finite horizon context.

We also bring to bear techniques from stochastic packing literature,
particularly the work on adaptivity gaps by Dean, Goemans and
Vondr\'ak~\cite{DeanGV04,DeanGV05,DeanThesis}.  Their results can be
viewed as sequential strategies for 2-level state spaces and is similar to the
online nature of the policies considered in stochastic
scheduling~\cite{MohringSU,SkutellaU}, where there is a strong notion of
``irrevocable commitment''. While the online notion is related to sequential
strategies, they are not the same.

In terms of analysis, our results can be thought of as extending analysis
both to arbitrary state spaces as well as for non-adaptive strategies for
the $2$-level case. Our overall technique can be thought of as ``LP
rounding via stochastic packing'' -- finding this connection between
finite horizon multi-armed bandits and stochastic packing by designing
simple LP rounding policies for a very general class of budgeted bandit
problems represents the key contribution of this work.

\paragraph{Related Work:}
Several heuristics had been proposed for the budgeted (futuristic) bandit
problem by Schneider and Moore~\cite{moore} and Madani \etal
\cite{madani}.  The final algorithm that arises from our framework bears
resemblance (but is not the same) to the algorithms proposed therein, but
as far as we are aware there was no prior analysis of any algorithm in
this context. A series of papers \cite{GoelGM06,Krause,GuhaM07} considered
the $2$-level state spaces (where a single play resolves all information
about an arm) for specific problems and presented approximations. The
Lagrangean (futuristic) bandit problem with $2$-level state space has been
considered before in~\cite{techrep}, where a $1.25$ approximation is
presented.  None of those techniques apply for the iterative refinement
that is required for multiple level state spaces.  Note that most other
literature on stochastic packing do not consider refinement of
information~\cite{KleinbergRT97,GoelI99}.

Our LP relaxation is well-studied in the context of multi-armed bandit
problems~\cite{BV06,whittle2,nino2} and other loosely coupled systems such
as multi-class queueing systems~\cite{gamarnik,lpbased}; we present the
first provable analysis of this formulation. Though LP formulations over
the state space of outcomes exist for other stochastic optimization
problems such as multi-stage optimization with
recourse~\cite{saa,ShmoysS04,CharikarCP05}, these formulations are based
on sampling scenarios. However these problems also do not have a notion of
refinement, and are fundamentally different from our setting where the
scenarios would be refinement trajectories~\cite{traject} that are hard to
sample.

\section{Types of State Spaces}
\label{sec:2level}
Recall that each arm is associated with a state that evolves when the arm is played. The state captures the distributional knowledge about the reward distribution of the arm. Formally, the set of possible states of arm $i$ is denoted by $\S_i$, and the initial state is $\rho_i \in \S_i$.  When the arm $i$ is played in a state $u \in \S_i$, the arm transitions to state $v \in \S_i$ w.p. $\p_{uv}$ depending on the observed outcome of the play.  The initial state models the prior knowledge about the arm. The states in general capture the posterior conditioned on the observations from a sequence of plays (or experiments) starting at the root.  Every state $u \in \S_i$ is associated with a reward $r_u$, which is the expected reward of playing in this state (which is of course conditioned on the observations from the plays so far).  By Bayes' rule, the reward of the different states evolve according to a {\em Martingale} property: $r_u = \sum_{v \in \S_i} \p_{uv} r_v$.

We now present two representative scenarios in order to better motivate
the abstract problem formulation. In the first scenario, the underlying
reward distribution is deterministic, and the distributional knowledge is
specified as a distribution over the possible deterministic values; this
implies that the uncertainty about an arm is completely resolved in one
play by observing the reward. In the second scenario, the uncertainty
resolves gradually over time.

\paragraph{Two-level State Space.}  A two-level state space models the case where the underlying reward of the arm is deterministic, so that the prior knowledge is a distribution over these values. In this setting, a single play resolves this distribution into a deterministic posterior. Formally, the prior distributional knowledge $X_i$ is a discrete distribution over values  $\{a^i_1,a^i_2,\ldots,a^i_m\}$, so that $\Pr[X_i = a^i_j] = p^i_j$ for $j = 1,2, \ldots,m$. The state space $\S_i$ of the arm is as follows: The root node $\rho_i$ has $r_{\rho_i} = \E[X_i] = \mu_i$.  For $j = 1,2, \ldots,m$, state $i_j$ has $r_{i_j} = a^i_j$, and $\p_{\rho_i i_j} = p^i_j$. Since the underlying reward distribution is simply a deterministic value, the  state space is $2$-level,  defining a star graph with $\rho_i$ being the root, and $i_1,i_2,\ldots,i_m$ being the leaves. 

To motivate budgeted bandits in such state spaces, consider a sensor network where the root server monitors the maximum value~\cite{BabcockO04,SilbersteinBEMY06}. The probability distributions of the values at various nodes are known to the server via past observations.  However, at the current step, probing all nodes to find out their actual values is undesirable since it requires transmissions from all nodes, consuming their battery life. Consider the simple setting where the network connecting the nodes to the server is a one-level tree, and probing a node consumes battery power of that node. Given a bound on the total battery life consumed, the goal of the root server is to maximize (in expectation) its estimate of the maximum value. Formally, each node corresponds to a distribution $X_i$ with mean $\mu_i$; the exact value sensed at the node can be found by paying a ``transmission cost"  $c_i$. The goal of the server is to adaptively probe a subset $S$ of  nodes with total transmission cost at most $C$ in order to maximize the estimate of the largest value sensed, {\em i.e} maximize $\E[\max \left( \max_{i \in S} X_i, \max_{i \notin S} \mu_i \right)]$, where the expectation is over the adaptive choice of $S$ and the outcome of the probes. The term $\max_{i \notin S} \mu_i $ incorporates the mean of the unprobed nodes into the estimate of the maximum value.

\medskip
In this context, it is desirable for the sensor node to probe the nodes in parallel, {\em i.e.}, use a {\bf non-adaptive strategy}. The question then becomes how good is such a strategy compared to the optimal adaptive strategy. We show positive results for the context of $2$-level spaces in Section~\ref{sec:super}.

\paragraph{Multi-level State Spaces.} These are the most general state spaces we consider, and make sense in contexts such as clinical trials where the underlying effectiveness of a treatment is a random variable following a parametrized distribution with unknown parameters. The prior distribution will then be a distribution over possible parameter values. In the clinical trial setting, each experimental drug is a bandit arm, and the goal is to devise a clinical trial phase to maximize the belief about the effectiveness of the drug finally chosen for marketing. Each drug has an effectiveness that is unknown a priori. The effectiveness can be modeled as a coin whose bias, $\theta$, is unknown a priori -- the outcomes of tossing the coin (running a trial) are $0$ and $1$ which correspond to a trial being ineffective and effective respectively. The uncertainty in the bias is specified by a {\em prior} distribution (or belief) on the possible values it can take.  Since the underlying distribution is Bernoulli, its conjugate prior is the Beta distribution.  A Beta distribution with parameters $\alpha_1,\alpha_2 \in \{1,2,\ldots\}$, which we denote $B(\alpha_1,\alpha_2)$ has p.d.f. of the form $c\theta^{\alpha_1-1} (1-\theta)^{\alpha_2-1}$, where $c$ is  a normalizing constant. $B(1,1)$ is the uniform distribution, which corresponds to having no a priori information.  The distribution $B(\alpha_1,\alpha_2)$ corresponds to the current (posterior) distribution over the possible values of the bias $\theta$ after having observed $(\alpha_1-1)$ $0$'s and $(\alpha_2-1)$ $1$'s. Given this distribution as our belief, the expected value of the bias or effectiveness is $\frac{\alpha_1}{\alpha_1+\alpha_2}$.

The state space $\S_i$ is a DAG, whose root $\rho_i$ encodes the initial belief about the bias, $B(\alpha_{1},\alpha_{2})$, so that $r_{\rho_i} = \frac{\alpha_1}{\alpha^{\rho}_1+\alpha_2}$. When the arm is played in this state, the state evolves depending on the outcome observed -- if the outcome is $1$, which happens w.p. $ \frac{\alpha_1}{\alpha_1+\alpha_2}$, the  child $u$ has belief  $B(\alpha+1,\alpha_2)$, so that $r_u = \frac{\alpha_1+1}{\alpha_1+\alpha_2+1}$, and $\p_{\rho u} = \frac{\alpha_1}{\alpha_1+\alpha_2}$; if the outcome is $0$, the child $v$ has belief $B(\alpha_1,\alpha_2+1)$, $r_v = \frac{\alpha_1}{\alpha_1+\alpha_2+1}$, and $\p_{\rho v} = \frac{\alpha_2}{\alpha_1+\alpha_2}$.  In general, if the DAG $\S_i$ has depth $C$ (corresponding to playing the arm at most $C$ times), it has $O(C^2)$ states. We omit details, since Beta distributions and their multinomial generalizations, the Dirichlet distributions, are standard in the Bayesian context (refer for instance Wetherill and Glazebrook~\cite{WG}).

\section{Budgeted Bandits}
\label{sec:budget}
We are given a bandit with $n$ independent arms. The set of possible
states of arm $i$ is denoted by $\S_i$, and the initial state is $\rho_i
\in \S_i$.  When the arm $i$ is played in a state $u \in \S_i$, the arm
transitions to state $v \in \S_i$ w.p. $\p_{uv}$. The reward at a state
satisfies $r_u = \sum_{v \in \S_i} \p_{uv} r_v$.
The cost of
a play depends on whether the previous play was for the same arm or not.
If the previous play was for the same arm, the play at $u \in \S_i$ costs
$c_u$, else it costs $c_u + h_i$, where $h_i$ is the setup cost for
switching into arm $i$. 
A policy $\pi$ performs a possibly adaptive sequence of plays
during the exploration.  leading to a probability distribution over
outcomes, $\O(\pi)$.  In outcome $o \in \O(\pi)$, each arm $i$ is in some
final state $u_i^o$. In this outcome $o$ the policy chooses $\max_i
r_{u^o_i}$.  The expected reward of the policy $\pi$ over the outcomes of
exploration, $R(\pi)$ is $\sum_{o \in \O(\pi)} q(o,\pi) \max_i r_{u^o_i}$.
Let $C(o,\pi)$ denote the cost of the exploration plays made by the policy
given an outcome $o$. In this section, we seek to find the policy $\pi$
which maximizes $\R(\pi)$ subject to $C(o,\pi) \leq C$ for all $o \in \O$.

We describe the linear programming formulation and rounding technique that
yields a $4$-approximation. We note that the formulation and solution are
polynomial in $n$, the number of arms, and $m$, the number of states per
arm. 
\subsection{Linear Programming Formulation}
\label{sec:lpsolve}
Recall the notation from Section~\ref{sec:probs}. Consider any adaptive
policy $\pi$. For some arm $i$ and state $u \in \S_i$, let: (1) $w_u$
denote the probability that during the execution of the policy $\pi$, arm
$i$ enters state $u \in \S_i$; (2) $z_u$ denote the probability that the
state of arm $i$ is $u$ and the policy plays arm $i$ in this state; and
(3) $x_u$ denote the probability that the policy $\pi$ chooses the arm $i$
in state $u$ during the exploitation phase. Note that since the latter two
correspond to mutually exclusive events, we have $x_u + z_u \le w_u$.  The
following LP which has three variables $w_u,x_u,$ and $z_u$ for each arm
$i$ and each $u \in \S_i$. A similar LP formulation was proposed for the
multi-armed bandit problem by Whittle~\cite{whittle2} and Bertsimas and
Nino-Mora~\cite{nino}.

\[ \mbox{Maximize}\ \  \sum_{i=1}^n \sum_{u \in \S_i} x_u r_u \]
\[ \begin{array}{rcll}
\sum_{i=1}^n  \left(h_i z_{\rho_i} +  \sum_{u \in \S_i} c_u z_u \right)& \le & C &\\
\sum_{i=1}^n \sum_{u \in \S_i} x_u & \le & 1\\
 \sum_{v \in \S_i} z_v \p_{vu} & = & w_u & \forall i,  u \in \S_i \setminus \{\rho_i\} \\
x_u + z_u & \le & w_u & \forall u \in \S_i, \forall i\\
x_u , z_u, w_u & \in & [0,1] & \forall u \in \S_i, \forall i\\
\end{array}\]
Let $\lpopt$ be the optimal LP value, and $OPT$ be the expected reward of the optimal adaptive policy.

\begin{claim}
\label{lem:key}
$OPT \le   \lpopt$.
\end{claim} 

\begin{proof} 
We show that the $w_u,z_u,x_u$ as defined above, 
corresponding to the optimal policy $\pi^*$, are feasible for the 
constraints of the LP.  Since each possible outcome of exploration 
leads to choosing one arm $i$ in some state $u \in \S_i$ for exploitation, 
in expectation over the outcomes, one arm in one state is chosen for 
exploitation. This is captured by the first constraint. Further, since on
each sequence of outcomes (the decision trajectory), the cost of playing
and switching into the arm is at most $C$, over the entire decision tree,
the expected cost of switching into the root states $\rho_i$ plus the
expected cost of play is at most $C$. This is captured by the second
constraint. Note that the LP only takes into account the cost of switching
into an arm the very first time this arm is explored, and ignores the rest
of the switching costs. This is clearly a relaxation, though the optimal
policy might switch multiple times into any arm. However, our rounding
procedure switches into an arm at most once, preserving the structure of
the LP relaxation. 

The third constraint simply encodes that the
probability of reaching a state $u \in \S_i$ during exploration.
It is precisely the probability with which it is played in some state $v \in \S_i$, times the probability $\p_{vu}$ that it reaches $u$ conditioned on that play. The constraint $x_u + z_u \le w_u$ simply captures that playing an arm is a disjoint event from exploiting it in any state. The objective is precisely the expected reward of the policy. Hence, the LP is a relaxation of the optimal policy.
\end{proof} 

\subsection{The Single-arm Policies}
\label{sec:explore}

\newcommand{\K}{\mathcal{K}}
\newcommand{\D}{\mathcal{D}} 
\renewcommand{\T}{\mathcal{S}} 
\newcommand{\h}{h}
\newcommand{\Eta}{\mathcal{E}}
\newcommand{\Z}{\mathcal{Z}}
\newcommand{\Y}{\mathcal{Y}}
\renewcommand{\L}{\mathcal{L}}
\renewcommand{\A}{\mathcal{\phi}}

The optimal LP solution clearly does not directly correspond to a feasible
policy since the variables do not faithfully capture the joint evolution
of the states of different arms. Below, we present an interpretation of
the LP solution, and show how it can be converted to a feasible
approximately optimal policy.

Let $\langle w^*_u,x^*_u,z^*_u \rangle$ denote the optimal solution to the
LP. We can assume w.l.o.g. that $w^*_{\rho_i} = 1$ for all $i$. Ignoring the
first two constraints of the LP for the time being, the remaining
constraints encode a separate policy for each arm as follows: Consider any
arm $i$ in isolation. The play starts at state $\rho_i$. The arm is played
with probability $z^*_{\rho_i}$, so that state $u \in \T_i$ is reached
with probability $z^*_{\rho_i} \p_{\rho_i u}$. This play incurs cost $h_i
+ c_{\rho_i}$, which captures the cost of switching into this arm, and the
cost of playing at the root. At state $\rho_i$, with probability
$x^*_{\rho_i}$, the play stops and arm $i$ is chosen for exploitation. The
events involving playing the arm and choosing for exploitation are
disjoint. Similarly, conditioned on reaching state $u \in \T_i$, with
probabilities $z^*_u/w^*_u$ and $x^*_u/w^*_u$, arm $i$ is played and
chosen for exploitation respectively. This yields a policy $\A_i$ for arm
$i$ which is described in Figure~\ref{fig1}. For policy $\A_i$, it is easy
to see by induction that if state $u \in \T_i$ is reached by the policy
with probability $w^*_u$, then state $u \in \T_i$ is reached {\em and} arm
$i$ is played with probability $z^*_u$.

The policy $\A_i$ sets $\Eta_i = 1$ if on termination, arm $i$ was chosen
for exploitation. If $\Eta_i = 1$ at state $u \in \T_i$, then exploiting
the arm in this state yields reward $r_u$.  Note that $\Eta_i$ is a random
variable that depends on the execution of policy $\A_i$. Let $R_i, C_i$
denote the random variables corresponding to the exploitation reward, and
cost of playing and switching, respectively.

\begin{figure*}[htbp]
\framebox{
\begin{minipage}{6.0in}
{\bf Policy $\A_i$:}
If arm $i$ is currently in state $u$, then choose $q \in [0,w^*_u]$ uniformly at random:
\begin{tabbing}
1.\ \ \= If $q \in [0,z^*_u]$, then play the arm  ({\bf explore}). \\
2.\> If $q \in (z^*_u,z^*_u+x^*_u]$, then stop executing $\A_i$, set  $\Eta_i=1$  ({\bf exploit}).  \\
3.\> If $q \in (z^*_u+x^*_u,w^*_u]$, then stop executing $\A_i$, set $\Eta_i =0$.
\end{tabbing}
\end{minipage}
}
\caption{\label{fig1} The Policy $\A_i$.}
\end{figure*}

For policy $\A_i$, define the following quantities:
\begin{enumerate}
\item $P(\A_i) = \E[\Eta_i] = \sum_{u \in \S_i}  \Pr[\Eta_i = 1 \wedge u] = \sum_{u \in \S_i} x^*_u$: Probability the arm is exploited. 
\item $R(\A_i) = \E[R_i] = \sum_{u \in \S_i} r_u \Pr[\Eta_i = 1 \wedge u] = \sum_{u \in \S_i} x^*_u r_u$: Expected reward of exploitation. 
\item $C(\A_i) = \E[C_i] = h_i z^*_i +  \sum_{u \in \S_i} c_u z^*_u$: Expected cost of switching into and playing this arm.
\end{enumerate}

Let $\A$ denote the policy that is obtained by executing each $\A_i$
independently in succession. Since policy $\A_i$ is obtained by
considering arm $i$ in isolation, $\A$ is {\bf not a feasible policy} for
the following reasons: (i) The cost $\sum_i C_i$ spent exploring all the
arms need not be at most $C$ in every exploration trajectory, and (ii) It
could happen that for several arms $i$, $\Eta_i$ is set to $1$, which
implies several arms could be chosen simultaneously for exploitation.

However, all is not lost. First note that the r.v. $R_i,C_i,\Eta_i$ for
different $i$ are independent. Furthermore, it is easy to see using the
first two constraints and objective of the LP formulation that $\A$ is
feasible in the following expected sense: $\sum_i \E[C_i] = \sum_i
C(\phi_i) \le C$.  Secondly, $\sum_i \E[\Eta_i] = \sum_i P(\phi_i) \le 1$.
Finally, $\sum_i \E[R_i] = \sum_i R(\phi_i) = \lpopt$.

Based on the above, we show that policy $\A$ can be converted to a
feasible policy using ideas from the adaptivity gap proofs for stochastic
packing problems~\cite{DeanGV04,DeanGV05,DeanThesis}.  We treat each
policy $\A_i$ as an item which takes up cost $C_i$, has size $\Eta_i$, and
profit $R_i$. These items need to be placed in a knapsack -- placing item
$i$ corresponds to exploring arm $i$ according to policy $\A_i$. This
placement is an irrevocable decision, and after the placement, the values
of $C_i, \Eta_i, R_i$ are revealed. We need $\sum_i C_i$ for items placed
so far should be at most $C$. Furthermore, the placement (or exploration)
stops the first time some $\Eta_i$ is set to $1$, and uses arm $i$ is used
for exploitation (obtaining reward or profit $R_i$). Since only one
$\Eta_i=1$ event is allowed before the play stops, this yields the "size
constraint" $\sum_i \Eta_i \le 1$. The knapsack therefore has both cost
and size constraints, and the goal is to sequentially and irrevocably
place the items in the knapsack, stopping when the constraints would be
violated.  The goal is to choose the order to place the items in order to
maximize the expected profit, or the exploitation gain. This is a
two-constraint stochastic packing problem. The LP solution implies that
the expected values of the random variables satisfy the packing
constraints.

We show that the ``start-deadline'' framework in~\cite{DeanThesis} can be
adapted to show that there is a fixed order of exploring the arms
according to the $\A_i$ which yields gain at least $\lpopt/4$.  There is
one subtle point -- the profit (or gain) is also a random variable
correlated with the size and cost. Furthermore, the ``start deadline''
model in~\cite{DeanThesis} would also imply the final packing could
violate the constraints by a small amount.  We get around this difficulty
by presenting an algorithm {\sc GreedyOrder} that explicitly obeys the
constraints, but whose analysis will be coupled with the analysis of a
simpler policy {\sc GreedyViolate} which exceeds the budget. The central
idea would be that although the benefit of the current arm has not been
``verified'', the alternatives have been ruled out.

\subsection{The Rounding Algorithm} 
\label{sec:greedy}
\begin{figure*}[htbp]
\framebox{
\begin{minipage}{6.0in}
{\bf Algorithm} \ {\sc GreedyOrder}
\begin{enumerate}\parskip=0pt
\item Order the arms in decreasing order of $\frac{R(\phi_i)}{P(\phi_i) + \frac{C(\phi_i)}{C}}$ and choose the arms to play in this order. 
\item \label{stepchange} For each arm $j$ in sorted order, play arm $j$ according to $\A_j$ as follows until $\A_j$ terminates:
\begin{enumerate}
\item \label{stepfirst} If the next play according to $\A_j$ would violate the budget constraint, then {\bf stop} exploration and {\bf goto} step (\ref{stepomit}). 
\item If $\A_j$ has terminated and $\Eta_j=1$, then {\bf stop} exploration and {\bf goto} step (\ref{stepomit}). 
\item Else, play arm $j$ according to policy $\A_j$ and {\bf goto} step (\ref{stepfirst}).
\end{enumerate}
\item \label{stepomit} Choose the {\em last arm} played in step (\ref{stepchange}) for exploitation.
\end{enumerate}
\end{minipage}
}
\caption{\label{fig2} The {\sc GreedyOrder} policy.}
\end{figure*}

The {\sc GreedyOrder} policy is shown in Figure~\ref{fig2}. Note that step (\ref{stepomit}) ensures that no arm is ever revisited, so that the strategy is {\em sequential}. For the purpose of analysis, we first present an infeasible policy {\sc GreedyViolate} which is simpler to analyze.  The algorithm is the same as {\sc GreedyOrder} except for step  (\ref{stepchange}), which we outline in Figure~\ref{fig3}. 

\begin{figure*}
\framebox{
\begin{minipage}{6in}
Step~\ref{stepchange} ({\sc GreedyViolate})  For each arm $j$ in sorted order, do the following:
\begin{enumerate}
\item[(a)] Play arm $j$ according to policy $\A_j$ until $\A_j$ terminates. 
\item[(b)] When the policy $\A_j$ terminates execution, if event $\Eta_j=1$ is
  observed or the cost budget $C$ is exhausted or exceeded, then {\bf
    stop} exploration and {\bf goto} step (\ref{stepomit}). 
\end{enumerate}
\end{minipage}
}
\caption{\label{fig3} The {\sc GreedyViolate} policy.}
\end{figure*}

In {\sc GreedyViolate}, the cost budget is checked only {\em after} fully executing a policy $\A_j$. Therefore, the policy could violate the budget constraint by at most the  exploration cost $c_{\max}$ of one arm.  

\begin{theorem}
\label{thm:violate}
{\sc GreedyViolate}  spends cost at most $C+c_{\max}$ and yields reward at least $\frac{OPT}{4}$.
\end{theorem}
\begin{proof}
  We have $\gamma^* = \sum_i R(\phi_i)$, and $\sum_i P(\phi_i) \le 1$.  We note that the random variables corresponding to different $i$ are independent.
  
For notational convenience, let $\nu_i = R(\phi_i)$, and let $\mu_i = P(\phi_i) +  C(\phi_i)/C$.  We therefore have $\sum_i \mu_i \le 2$.  The sorted ordering is decreasing order of $\nu_i/\mu_i$. Re-number the arms  according to the sorted ordering so that the first arm played is  numbered $1$. Let $k$ denote the smallest integer such that $ \sum_{i=1}^k \mu_i \ge 1$.  By the sorted ordering property, it is easy to see that $\sum_{i=1}^k  \nu_i \ge  \frac{1}{2} \gamma^*$.  
 
 Arm $i$ is reached and played by the policy iff $\sum_{j<i} \Eta_j = 0$, and $\sum_{j < i} C_j < C$. This translates to $\sum_{j < i} \left( \Eta_j + \frac{C_j}{C} \right) < 1$.  Note that $\E[\Eta_j + \frac{C_j}{C}] = P(\phi_j) + C(\phi_j)/C = \mu_j$. Therefore, by Markov's inequality, $\Pr \left[\sum_{j < i} \left( \Eta_j + \frac{C_j}{C} \right) < 1 \right] \ge \max(0,1 - \sum_{j < i} \mu_j)$. Note further that for  $i \leq k$, we have $\mu_i \le 1$.  
 
 If arm $i$ is played,  it yields  reward $\nu_i$ that directly contributes to the exploitation reward. Since the reward is independent of the event that the arm is reached and played. Therefore, the expected reward of {\sc GreedyViolate} can be bounded by
linearity of expectation as follows.
\[ \mbox{Reward of {\sc GreedyViolate}= } \G \ge  \sum_{i =1}^k (1- \sum_{j < i} \mu_j) \nu_i \]
We now follow the proof idea in~\cite{DeanThesis}. Consider the arms $1 \le i \le k$ as deterministic items with item $i$ having profit $\nu_i$ and size $\mu_i$. We therefore have $\sum_{i =1}^k \nu_i \ge \gamma^*/2$ and $\sum_{i=1}^{k-1} \mu_i  \le 1$.

Suppose these items are placed into a knapsack of size $1$ in decreasing order of $\frac{\nu_i}{\mu_i}$ with the last item possibly being fractionally placed. This is the same ordering that the algorithm uses to play the arms.  Let $\Phi(q)$ denote the profit when size of the knapsack filled is $q \le 1$. We have $\Phi(1) \ge \gamma^*/2$. Plot the function $\Phi(q)$ as a function of $q$. This plot connects the points $\{(0,0), (\mu_1,v_1),(\mu_1+\mu_2,v_1+v_2), \ldots (1,\Phi(1))\}$. This function is concave, therefore the area under the curve is at least $\frac{\Phi(1)}{2} \ge \gamma^*/4$. However, the area under this curve is at most  
$$v_1 + v_2 (1-\mu_1)  + \ldots + v_k (1-\sum_{j < k} \mu_j) \le \G$$ 
Therefore, $\G \ge \gamma^*/4$. Since $OPT \le \gamma^*$, $\G$ is at least $\frac{OPT}{4}$.
\end{proof}


\begin{theorem}
\label{thm:order}
The {\sc GreedyOrder} policy with cost budget $C$ achieves reward at least
$\frac{OPT}{4}$.
\end{theorem}
\begin{proof}
  Consider the {\sc GreedyViolate} policy. This policy could exceed the  cost budget because the budget was checked only at the end of execution  of policy $\A_i$ for arm $i$. Now suppose the play for arm $i$ reaches  state $u \in \H_i$, and the next decision of {\sc GreedyViolate}  involves playing arm $i$ and this would exceed the cost budget. The {\sc  GreedyViolate} policy continues to play arm $i$ according to $\A_i$  and when the play is finished, it checks the budget constraint, realizes that the budget is exhausted, stops, and chooses arm $i$ for  exploitation. Suppose the policy was modified so that instead of the  decision to play arm $i$ further at state $u$, the policy instead checks  the budget, realizes it is not sufficient for the next play, stops, and  chooses arm $i$ for exploitation.  This new policy is precisely {\sc GreedyOrder}.

Note now that conditioned on reaching node $u$ with the next decision of {\sc GreedyViolate} being to play arm $i$, so that the policies {\sc GreedyViolate} and {\sc GreedyOrder} diverge in their next action, both policies choose arm $i$ for exploitation. By the martingale property of the rewards, the reward from choosing arm $i$ for exploitation at state $u$ is the same as the expected reward from playing the arm further and then choosing it for exploitation. Therefore, the expected reward of both policies is identical, and the theorem follows.
\end{proof}

\subsection{Bi-criteria Result}
Suppose we allow the cost budget to be exceeded by a factor $\alpha \ge 1$, so that the cost budget is $\alpha C$. Consider the {\sc GreedyOrder} policy where the arms are ordered in decreasing order of $\frac{R(\phi_i)}{\alpha P(\phi_i) +  C(\phi_i)/C}$, and the budget constraint is relaxed to $\alpha C$. We have the following theorem:

\begin{theorem}
For any $\alpha \ge 1$, if the cost budget is relaxed to $\alpha C$, the expected reward of the modified {\sc GreedyOrder} policy is $\frac{\alpha}{2(1+\alpha)} \gamma^*$.
\end{theorem}
\begin{proof}
We mimic  the proof of Theorem~\ref{thm:violate},  and define $\nu_i = R(\phi_i)$, and let $\mu_i = P(\phi_i) +  \frac{1}{\alpha} C(\phi_i)/C$.  Note that the LP satisfies the constraint $\sum_i \left(P(\phi_i) + \frac{1}{\alpha} \frac{C(\phi_i)}{C} \right) \le \frac{1+\alpha}{\alpha}$. We therefore have $\sum_i \mu_i \le \frac{1+\alpha}{\alpha}$.  Let $k$ denote the smallest integer such that $ \sum_{i=1}^k \mu_i \ge 1$.  By the sorted ordering property, we have $\sum_{i=1}^k  \nu_i \ge  \frac{\alpha}{1+\alpha} \gamma^*$.  The rest of the proof remains the same, and we show that the reward of the new policy, $\G$, satisfies: $ \G \ge \frac{1}{2} \Phi(1)$, and $\Phi(1) \ge   \frac{\alpha}{2(1+\alpha)} \gamma^*$. This completes the proof.
\end{proof}

\subsection{Integrality Gap of the Linear Program}
We now show via a simple example that the linear program has an integrality gap of at least $e/(e-1) \approx 1.58$.  All arms $i=1,2,\ldots,n$ have identical $2$-level state spaces. Each $\S_i$ has $c_{\rho} = 1$, $r_{\rho} = 1/n$, switching cost $h_i = 0$, and two other states $u_0$ and $u_1$. We have $\p_{\rho u_0} = 1-1/n$, $\p_{\rho u_1} = 1/n$, $r_{u_0} = 0$, $r_{u_1} = 1$.  Set  $C = n$, so that any policy can play all the arms. The expected reward of such a policy is precisely $1 - (1-1/n)^n \approx 1 - 1/e$. The LP solution will set $z^*_{\rho} = 1$  and $x^*_{u_1} = 1/n$ for all $i$, yielding an LP objective of $1$. This shows that the linear program cannot yield better than a constant factor approximation. It is an interesting open question whether the LP can be strengthened by other convex constraints to obtain tighter bounds (refer for instance~\cite{DeanThesis}).

\section{Non-adaptive Policies: Bounding the Adaptivity Gap}
\label{sec:super}
\label{sec:nonadapt}
Recall  that a non-adaptive strategy allocates a fixed budget to each arm in advance. It then explores the arms according to these budgets (ignoring the outcome of the plays in choosing the next arm to explore), and at the end of exploration, chooses the best arm for exploitation.  This is termed an {\em allocational strategy} in~\cite{madani}. Such strategies are desirable since they allow the experimenter to consider various competing arms in parallel. We show two results in this case: For general state spaces, we show that such a non-adaptive strategy can be arbitrarily worse than the optimal adaptive strategy. On the positive side, we show that for $2$-level state spaces, which correspond to deterministic underlying rewards (refer Section~\ref{sec:2level}), a non-adaptive strategy is only a factor $7$ worse than the performance of the optimal adaptive strategy.

\subsection{Lower Bound for Multi-level State Spaces}
We first present an example with unit costs where an adaptive strategy that dynamically allocates the budget achieves far better exploitation gain than a non-adaptive strategy. Note that we can ignore switching costs in such strategies.
\begin{theorem}
The adaptivity gap of the budgeted learning problem is $\Omega(\sqrt{n})$. Furthermore, even if we allow the non-adaptive exploration to use $\gamma>1$ times the exploration budget, the adaptivity gap remains $\Omega(\sqrt{n/\gamma})$.
\end{theorem}
\begin{proof}
 Each arm has an underlying reward distribution over the three values $a_1 = 0$, $a_2 = 1/n^{9}$ and $a_3 = 1$. Let $q =  1/\sqrt{n}$.  The underlying distribution could be one of $3$ possibilities:  $R_1, R_2, R_3$. $R_1$ is the deterministic value $a_1$,
  $R_2$ is deterministically $a_2$ and $R_3$ is $a_3$ w.p. $q$ and $a_2$  w.p. $1-q$. For each arm, we know in advance that $\Pr[R_1] = 1-q$, $\Pr[R_2] =  q(1-q)$ and $\Pr[R_3] = q^2$. Therefore, the knowledge for each arm is a prior over the three distributions $R_1, R_2,R_3$. The priors for different arms are i.i.d. All $c_i = 1$ and the total budget is $C =  5n$.

 We first show that the adaptive policy chooses an arm with underlying reward distribution $R_3$ with constant probability. This policy first  plays each arm once and discards all arms with observed reward $a_1$. With probability at least  $1/2$, there are at most $2/q$ arms which survive,  and at least one of these arms has underlying reward distribution $R_3$. If more arms survive, choose any $2/q$ arms. The policy now plays each of the $2/q$ arms $2\sqrt{n}$ times. The  probability that an arm with distribution $R_3$ yields reward $a_3$ on some play is at  least once is $1-(1-q)^{2/q} \approx \Theta(1)$. In this case, it  chooses the arm with reward distribution $R_3$ for exploitation. Since this happens w.p. at least a constant, the expected  exploitation reward is $\Theta(q)$. Note that this is best possible to within constant factors, since $\E[R_3] = \Theta(q)$.

Now consider any non-adaptive policy. With probability $1 - 1/n^{\Theta(1)}$, there are at most $ 2 \log n$ arms with  reward distribution $R_3$, and at least $1/(2q)$ arms with reward distribution $R_2$.   Let $r\gg 2\log n$. The strategy allocates at most $5r$  plays to at least $n(1-1/r)$ arms -- call this set of arms $T$. With  probability  $(1-1/r)^{2 \log n} =\Omega(1-(2\log n)/r)$, all arms with reward  distribution $R_3$ lie in this set $T$.  For any of these arms played $O(r)$ times, with probability $1-O(qr)$, all observed rewards will have  value $a_2$. This implies with probability $1-O(qr)$, all arms with distribution $R_3$ yield rewards $a_2$, and so do $\Omega(1/(2q))$ arms  with distributions $R_2$. Since these appear indistinguishable to the policy, it can at best choose one of these at random, obtaining exploitation reward $\frac{q \log n}{2(1/q)} = O(q^2 \log n)$. Since this situation happens  with probability $1-O(\log n/r)$, and with the remaining probability the  exploitation reward is at most $q$, the strategy therefore has expected exploitation reward $O(q\log n(\frac{1}{r} +q))$. This implies the adaptivity gap  is $\Omega(1/q)=\Omega(\sqrt{n})$ if we set $r=1/q$.

 Now suppose we allow the budget to be increased by a factor of $\gamma>1$. Then the strategy would allocate at most $5\gamma r$ plays  to at least $n(1-1/r)$ arms. By following the same argument as above,  the expected reward is $O(q\log n(\frac{1}{r} +q \gamma))$. This proves  the second part of the theorem.
\end{proof}

\subsection{Upper Bound for Two-Level State Spaces} 
We next show that for $2$-level state spaces, which correspond to deterministic underlying rewards (refer Section~\ref{sec:2level}), the adaptivity gap is at most a factor of $7$.
\begin{theorem}
If each state space $\S_i$ is a directed star graph with $\rho_i$ as the root, then there is a non-adaptive strategy that achieves reward at least $1/7$ the LP bound.
\end{theorem}
\begin{proof}
In the case of $2$-level state spaces, a non-adaptive strategy chooses a subset $S$ of arms and allocates zero/one plays to each of these so that the total cost of the plays is at most $C$. We consider two cases based on the LP optimal solution.

In the first case, suppose $\sum_i r_{\rho_i} x_{\rho_i} \ge \gamma^*/7$, then not playing anything but simply choosing the arm with highest $r_{\rho_i}$ directly for exploitation is a $7$-approximation.

In the remaining proof, we assume the above is not the case, and compare against the optimal LP solution that sets $x_{\rho_i} = 0$ for all $i$. This solution has value at least $6 \gamma^*/7$. For simplicity of notation, define $z_i = z_{\rho_i}$ as the probability that the arm $i$ is played. Define  $X_i = \frac{1}{z_i} \sum_{u \in \S_i} x_u$ as the probability that the arm is exploited conditioned on being played, and $R_i = \frac{1}{z_i} \sum_{u \in S_i} x_u r_u$ as the expected exploitation reward conditioned on being played. Also define $c_i = c_{\rho_i}$. The LP satisfies the constraint:  $ \sum_i  z_i \left( \frac{c_i}{C} + X_i \right)  \le 2$, and the LP objective is $\sum_i z_i R_i$, which has value at least $6 \gamma^*/7$.

A better objective for the LP can be obtained by considering the arms in decreasing order of $\frac{R_i}{ \frac{c_i}{C} + X_i}$, and increasing $z_i$ in this order until the constraint $ \sum_i  z_i \left( \frac{c_i}{C} + X_i \right)  \le 1$ becomes tight. Set the remaining $z_i = 0$. It is easy to see  $\sum_i z_i R_i \ge \frac{3}{7} \gamma^*$.  At this point, let $k$ denote the index of the last arm which could possibly have $z_k < 1$, and let $S$ denote the set of arms with $z_i = 1$ for $i \in S$.  There are again two cases.

In the first case, if $z_k R_k > \gamma^*/7$, then choosing just this arm for exploitation has reward at least $\gamma^*/7$, and is a $7$-approximation. 

In the second and final case, we have a subset of arms $ \sum_{i \in S} \left( \frac{c_i}{C} + X_i \right)  \le 1$, and $\sum_{i \in S} R_i \ge  \frac{3}{7} \gamma^* - \gamma^*/7 = \frac{2}{7}\gamma^*$. If all these arms are played, the expected number of arms that are exploited is  $\sum_{i \in S} X_i \le 1$, and the expected reward is $\sum_{i \in S} R_i \ge \frac{2}{7}\gamma^*$. The proof of Theorem~\ref{thm:violate} can be adapted to show that choosing the best arm for exploitation yields at least half the reward, {\em i.e.}, reward at least $\gamma^*/7$.
\end{proof}

\section{Lagrangean Version}
\label{sec:lag}
Recall from Section~\ref{sec:probs} that in the Lagrangean version of the problem, there are no budget constraints on the plays, the goal is to find a policy $\pi$ such that $R(\pi) - C(\pi)$ is maximized. Denote this quantity as the {\em profit} of the strategy.

The linear program relaxation is below. The variables are identical to the previous formulation, but there is no budget constraint. 

\[ \mbox{Maximize}\ \  \sum_{i=1}^n \left( \sum_{u \in \H_i} \left(x_u r_u - c_u z_u \right) - h_i z_{\rho_i} \right) \ \]
\[ \begin{array}{rcll}
\sum_{i=1}^n \sum_{u \in \T_i} x_u & \le & 1\\
 \sum_{v \in \S_i} z_v \p_{vu} & = &w_u & \forall  i, u \in \H_i \setminus \{\rho_i\}  \\
x_u + z_u & \le & w_u & \forall u \in \H_i, \forall i\\
x_u , z_u, w_u & \in & [0,1] & \forall u \in \H_i, \forall i\\
\end{array}\]

\noindent Let $OPT =$  optimal net profit and $\lpopt=$ optimal LP solution. The next  is similar to Claim~\ref{lem:key}.
\begin{claim}
$OPT \le \lpopt$. 
\end{claim}

From this LP optimum $\langle w^*_u,x^*_u,z^*_u \rangle$,  the policy $\A_i$ is constructed as described in Figure~\ref{fig1}, and the r.v.'s  $\Eta_i,C_i,R_i$ and their respective expectations $P(\phi_i), C(\phi_i)$, and $R(\phi_i)$ are obtained as described in the beginning of Section~\ref{sec:explore}. Let r. v. $Y_i = R_i - C_i$ denote the  profit of playing arm $i$ according to $\A_i$. Note that $\E[Y_i] =  \left( \sum_{u \in \H_i} \left(x_u r_u - c_u z_u \right) - h_i z_{\rho_i} \right) $.

The nice aspect of the proof of Theorem~\ref{thm:violate} is that it does not necessarily require the r.v. corresponding to the reward of policy $\phi_i$, $R_i$ to be non-negative. As long as $\E[R_i] = R(\phi_i) \ge 0$, the proof holds. This will be crucial for the Lagrangean version.

\begin{claim} \label{claim:ge}
For any arm $i$, $\E[Y_i] = R(\phi_i) - C(\phi_i) \ge 0$.
\end{claim}
\begin{proof}
For each $i$, since all $r_u \ge 0$, setting $x_{\rho_i} \leftarrow \sum_{u \in \T_i} x_u$, $w_{\rho_i} \leftarrow 1$,  and $z_u \leftarrow 0$ for $u \in \T_i$ yields a feasible non-negative solution. The LP optimum will therefore guarantee that the term $\sum_{u \in \H_i}  \left(x_u r_u - c_u z_u \right) - h_i z_{\rho_i} \ge 0$. Therefore,  $\E[Y_i] \ge 0$ for all $i$.
\end{proof}

The {\sc GreedyOrder} policy orders the arms in decreasing order of $\frac{R(\phi_i) - C(\phi_i)}{P(\phi_i)}$, and plays them according to their respective $\A_i$ until some $\Eta_i=1$. 

\begin{theorem} \label{thm:lag}
The expected profit of {\sc GreedyOrder} is at least $OPT/2$.
\end{theorem}
\begin{proof}
Let $\mu_i = P(\phi_i)$ and $\nu_i = \E[Y_i]$ for notational convenience. The LP solution yields $\sum_i \mu_i  \le 1$ and $\sum_i \nu_i  = \lpopt$. Re-number the arms   according to the sorted ordering of $\frac{\nu_i}{\mu_i}$ so that the first arm played is  numbered $1$.  
 
The event that {\sc GreedyOrder} plays arm $i$  corresponds to $\sum_{j < i} \Eta_j=0$. By Markov's inequality, we have $\Pr[\sum_{j < i} \Eta_j=0] =  \Pr[\sum_{j < i} \Eta_j< 1]\ge 1 - \sum_{j<i} \mu_j$. 

If arm $i$ is played,  it yields profit $Y_i$. This implies the profit of {\sc GreedyOrder} is   $\sum_i Y_i (1-\sum_{j < i} \Eta_j)$. Since $Y_i $ is independent of $\sum_{j < i} \Eta_j$, and since  Claim~\ref{claim:ge} implies $\E[Y_i] \ge 0$,  the expected profit $\G$ of {\sc GreedyOrder} can be bounded by linearity of expectation as follows.
$$ \G = \sum_i \Pr \left[\sum_{j < i} \Eta_j< 1 \right]  \E[Y_i] \ge  \sum_i \nu_i \left(1-\sum_{j < i} \mu_j \right) $$
We now follow the proof idea in~\cite{DeanThesis}. Consider the arms $1 \le i \le n$ as deterministic items with item $i$ having profit $\mu_i$ and size $\mu_i$. We therefore have $\sum_i \nu_i \ge \gamma^*$ and $\sum_i \mu_i \le 1$.  Using the same proof idea as in Theorem~\ref{thm:violate}, it is easy to see that $\G \ge \frac{\gamma^*}{2}$. Since $OPT \le \gamma^*$, $\G$ is at least $\frac{OPT}{2}$.
\end{proof}

\section{Concave Utility Functions}
\label{sec:exten}
\renewcommand{\ss}{\mathbf{s}} The above framework in fact solves the more
general problem of maximizing any concave stochastic objective function
over the rewards of the arms subject to a (deterministic) packing
constraint. Several such examples of concave objective function are given
in \cite{low} in the context of optimizing ``TCP friendly'' network utility functions.
 In what follows, we extend our arguments in the previous section to develop approximation algorithms for all positive concave utility maximization problems in this exploration-exploration setting. Suppose arm $i$ in state $u \in \S_i$ has a value function $g_u(y)$ where $y \in [0,1]$ denotes the weight assigned to it in the exploitation phase.  We enforce the following properties on the function $g_u(y)$: 
\begin{description}
\item[Concavity.] $g_u(y)$ is an arbitrary positive non-decreasing concave function of $y$.
\item[Super-Martingale.]  $g_u(y) \ge \sum_{v \in \S_i} \p_{uv} g_v(y)$.
\end{description}

Given an outcome $o \in \O(\pi)$ of exploration, suppose arm $i$ ends up in state $u$, and is assigned weight $y_i$  in the exploitation phase, the contribution of this arm to the exploitation value is $g_u(y_i)$. The assignment of weights is subject to a deterministic packing constraint $\sum_i \sigma_i y_i \le B$, where $\sigma_{i} \in [0,B]$. Therefore, for a given outcome $o \in \O(\pi)$, the value of this outcome is given by the convex program:
\[\max  \sum_{i=1}^n    g_u(y_i)  \qquad  \mbox{s.t.} \qquad \sum_{i=1}^n \sigma_i y_i  \le  B,\forall i \ \ y_i  \in  [0,1] \]
The goal as before is to design an adaptive exploration phase $\pi$ so that the expected exploitation value is maximized, where the expectation is over the outcomes $\O(\pi)$ of exploration and cost of exploration is at most $C$. 
\begin{itemize}
\item For the maximum reward problem, $g_u(y)=r_u y$,  $\sigma_i=1$, and $B=1$.
\item Suppose we wish to choose the $m$ best rewards, we simply set $B = m$.  Note that we can also conceive of a scenario where the $c_i$ correspond to cost of ``pilot studies'' and each treatment $i$ requires cost $\sigma_i$  for large scale studies. This would lead us to a {\sc   Knapsack} type problem where $\sigma_i$ are now the ``sizes''. 
\end{itemize}  

\subsection{Linear Program} 
The state space $\H_i$ and the probabilities $\p_{uv}$ are defined just as in Section~\ref{sec:probs}.  For small constant $\epsilon > 0$, let $L = \frac{n}{\epsilon}$. Discretize the domain $[0,1]$ in multiples of $1/L$.  For $l \in \{0,1,\ldots,L\}$, let $\zeta_u(l) = g_u(l/L)$. This corresponds to the contribution of arm $i$ to the exploitation value on allocating weight $y_i = l/L$. Define the following linear program:
\[ \mbox{Max}\ \  \sum_{i=1}^n \sum_{u \in \H_i} \sum_{l=0}^L x_{ul} \zeta_u(l) \]
\[ \begin{array}{rcl}
\sum_{i=1}^n \left( h_i z_{\rho_i} + \sum_{u \in \H_i}  c_u z_u \right) & \le & C \\
\sum_{i=1}^n \sigma_i \left(\sum_{u \in \H_i} \sum_{l=0}^L l x_{ul} \right) & \le & BL(1+\epsilon) \\
 \sum_{v:u \in D(v)} z_v \p_{vu} & = &w_u \ \  \forall  i, u \in \H_i \setminus \{\rho_i\} \\
z_u  + \sum_{l=0}^L x_{ul} & \le & w_u \ \  \forall u \in \H_i, \forall i\\
w_u,x_{ul},z_u & \in & [0,1] \ \  \forall u \in \H_i, \forall i,l\\
\end{array} \]

Let $\lpopt$ be the optimal $LP$ value and $OPT =$ value of the optimal adaptive exploration policy.

\begin{lemma}
$OPT \le \lpopt$.
\end{lemma}
\begin{proof}
  In the optimal solution, let $w_u$ denote the probability that the
  policy reaches state $u \in \T_i$, and let $z_u$ denote the probability
  of reaching state $u \in \T_i$ and playing arm $i$ in this state. For $l
  \ge 1$, let $x_{ul}$ denote the probability of stopping exploration at
  $u \in \T_i$ and allocating weight $y_i \in (\frac{l-1}{L},\frac{l}{L}]$
  to arm $i$.  All the constraints are straightforward, except the
  constraint involving $B$. Observe that if the weight assignments $y_i$
  in the optimal solution were rounded up to the nearest multiple of
  $1/L$, then the total size of any assignment increases by at most
  $\epsilon B$ since all $s_i \le B$.  Therefore, this constraint is
  satisfied.  Using the same rounding up argument, if the weight satisfies
  $y_i \in (\frac{l-1}{L},\frac{l}{L}]$, then the contribution of arm $i$
  to the exploitation value is upper bounded by $ \zeta_u(l)$ since the
  function $g_u(y)$ is non-decreasing in $y$. Therefore, the proof follows.
\end{proof}

\subsection{Exploration Policy}
Let $\langle w^*_u,x^*_{ul},z^*_u \rangle$ denote the optimal solution to the $LP$. Assume $w^*_{\rho_i} = 1$ for all $i$. Also w.l.o.g,  $z^*_u  + \sum_{l=0}^L x^*_{ul} = w^*_u$ for all $u \in \T_i$.  The LP solution yields a natural (infeasible) exploration policy $\A$ consisting of one independent policy $\A_i$ per arm $i$. Policy $\A_i$ is described in Figure~\ref{fig4}.

\begin{figure*}
\framebox{
\begin{minipage}{6.0in}
{\bf Policy $\A_i$:}
If arm $i$ is currently in state $u$, choose $q \in [0,w^*_u]$ u.a.r. and do one of the following:
\begin{tabbing}
\ \ \=1.\= \ \=If $q \in [0,z^*_u]$,  {\bf then} play the arm.\\
\>2.\> \> {\bf else} Stop executing $\A_i$. \\
\> \> \> Find the smallest $l \ge 0$ such that $q \le z^*_u+\sum_{k=0}^{l}x^*_{uk}$. Set  $\Eta_i=\frac{l}{L}$ and $R_i = \zeta_u(l)$.
\end{tabbing}
\end{minipage}
}
\caption{\label{fig4} The policy $\A_i$ for concave value functions.}
\end{figure*}

The policy $\A_i$ is independent of the states of the other arms. It is easy to see by induction that if state $u \in \T_i$ is reached by the policy with probability $w^*_u$, then state $u \in \T_i$ is reached {\em and} arm $i$ is played with probability $z^*_u$.  Let random variable $C_i$ denote the cost of executing $\A_i$, and let $C(\phi_i) = \E[C_i]$. Denote this overall policy $\A$ -- this corresponds to one independent decision policy $\A_i$ (determined by $\langle w^*_u,x^*_{ul},z^*_u\rangle$) per arm.  It is easy to see that the following hold for $\A$:
\begin{enumerate}
\item $C(\phi_i) = \E[C_i] = h_i z^*_{\rho_i} + \sum_{u \in \T_i} c_u z^*_u$  so that $\sum_i C(\phi_i) \le C$. 
\item $P(\phi_i) = \E[\Eta_i] = \frac{1}{L} \sum_{u \in \T_i} \sum_{l=0}^L l x^*_{ul} \ \ \Rightarrow \ \ \sum_i \sigma_i P(\phi_i) \le B(1+\epsilon)$.  
\item $R(\phi_i) = \E[R_i] = \sum_{u \in \T_i} \sum_{l=0}^L x^*_{ul} \zeta_u(l) \ \ \Rightarrow  \ \ \sum_i R(\phi_i) = \lpopt$.
\end{enumerate}

\begin{figure*}[htbp]
\framebox{
\begin{minipage}{6.0in}
{\bf Algorithm} \ {\sc GreedyOrder}
\begin{enumerate}
\item Order the arms in decreasing order of $\frac{R(\phi_i)}{\frac{\sigma_i}{B} P(\phi_i) + \frac{1}{C}C(\phi_i)}$. 
\item \label{stepchange2} For each arm $j$ in sorted order, play it according to $\A_j$ as follows until $\A_j$ terminates:
\begin{enumerate}
\item \label{stepfirst2} If the next play would violate the cost constraint, then set $\Eta_j \leftarrow 1$, {\bf stop} exploration, and {\bf goto} step (\ref{stepomit2}). 
\item If $\A_j$ terminates and $\sum_i \sigma_i \Eta_i \ge B$, then {\bf stop} exploration and {\bf goto} step (\ref{stepomit2}). 
\item Else, play arm $j$ according to policy $\A_j$ and {\bf goto} step (\ref{stepfirst2}).
\end{enumerate}
\item \label{stepomit2} {\bf Exploitation:} Scale down $\Eta_i$ by a factor of $2$.
\end{enumerate}
\end{minipage}
}
\caption{\label{fig5}The {\sc GreedyOrder} policy for concave functions.}
\end{figure*}

The  {\sc GreedyOrder} policy is presented in Figure~\ref{fig5}. We again use an infeasible policy {\sc GreedyViolate} which is simpler to analyze. The algorithm is the same as {\sc GreedyOrder} except for step  (\ref{stepchange2}), where violation of the cost  constraint is only checked after the policy $\A_j$ terminates.

\begin{theorem}
Let $c_{\max}$ denote the maximum cost of exploring a single arm. Then  {\sc GreedyViolate}  spends cost at most $C+c_{\max}$ and has expected value $\frac{OPT}{8}(1-\epsilon)$.
\end{theorem}
\begin{proof}
Let $\nu_i = R(\phi_i)$ and let $\mu_i =  \frac{\sigma_i}{B} P(\phi_i) + \frac{1}{C}C(\phi_i)$. The LP constraints imply that  $\gamma^* = \sum_i \nu_i$, and  $\sum_i \mu_i \le 2+\epsilon$.  Now using the same proof as Theorem~\ref{thm:violate}, we obtain the value $\G$ of {\sc GreedyViolate} according to the weight assignment $\Eta_i$  at the end of Step (\ref{stepchange2}) is at least $\frac{OPT}{4}(1-\epsilon)$. This weight assignment could be infeasible because of  the last arm, so that the $\Eta_i$ only satisfy $\sum_i \sigma_i \Eta_i \le 2B$. This is made feasible in Step (\ref{stepomit2}) by scaling all $\Eta_i$ down by a factor of $2$. Since the functions $g_i(y)$ are concave in $y$, the exploitation value reduces by a factor of $1/2$ because of scaling down.
\end{proof}

\begin{theorem} 
{\sc GreedyOrder} policy with budget $C$ achieves expected value at least $\frac{OPT}{8} (1-\epsilon)$.
\end{theorem}
\begin{proof}
 Consider the {\sc GreedyViolate} policy.  Now suppose the play for arm $i$ reaches  state $u \in \H_i$, and the next decision of {\sc GreedyViolate}  involves playing arm $i$ and this would exceed the cost budget.  Conditioned on this next decision,   {\sc GreedyOrder} sets $\Eta_i=1$ and stops exploration. In this case, the exploitation value of {\sc GreedyOrder} from arm $i$ is at least the expected exploitation gain of {\sc GreedyViolate} for this arm by the super-martingale property of the value function $g$. Therefore,  for the assignments at the end of Step (\ref{stepchange2}), the gain of {\sc GreedyOrder} is  at least $\frac{OPT}{4}(1-\epsilon)$. Since Step (\ref{stepomit2}) scales the $\Eta$'s down by a factor of $2$,  the theorem follows.
\end{proof}

\section{Conclusions}
We studied the classical stochastic multi-armed bandit problem under the
future utilization objective in the presence of priors. This model is
relevant to settings involving data acquisition and design of experiments.
In this problem the exploration phase necessarily precedes the
exploitation phase. This makes the problem significantly different from
the problems in online optimization, which seeks to minimize regret over
the past, because online optimization models problems where exploration
and exploitation are simultaneous. The central difficulty of online
optimization is the lack of information, whereas the difficulty in 
optimizing future utilization is computational.
In fact the latter is provably  {\sc
  NP-Hard}. We presented constant factor approximation algorithms that
yield sequential policies for several extensions of this basic problem.
These algorithms proceed via LP rounding and show a surprising connection
to stochastic packing algorithms. We also show that the sequential
policy we develop is within constant factor of a fully adaptive
solution. Note that a constant factor adaptivity gap result is the best possible.

There are several challenging open questions arising from this work; we
mention two of them. First, we conjecture that  constructing a (possibly adaptive) strategy for the budgeted learning
problem is {\sc APX-Hard}, {\em i.e.},  there exists an absolute constant $c>1$ such that it is {\sc NP-Hard} to produce a
solution which is within factor $c$ times the optimum. Secondly, we have
focused exclusively on utility maximization; it would be interesting to
explore other objectives, such as minimizing residual
information~\cite{Krause}.

\medskip
\noindent {\bf Acknowledgment:} We would like to thank Jen Burge, Vincent
Conitzer, Ashish Goel, Ronald Parr, and Fernando Pereira for helpful
discussions.

\end{document}